\documentclass[fleqn,10pt]{wlscirep}
\usepackage[utf8]{inputenc}
\usepackage[T1]{fontenc}
\usepackage{xcolor}
\usepackage{float}

\title{Deep Learning for Automated Quantification of Tumor-Associated Macrophages from H\&E-Stained Slides in Diffuse Large B-Cell Lymphoma}

\author[1,2,+]{Anastasiia Studenikina}
\author[1,2,+,*]{Svetlana Illarionova}
\author[3]{Joaquim Carreras}
\author[4]{Daniil Sulimov}
\author[5]{Olga Filimonova}   
\author[5,6]{Dmitry Zvezdin}
\author[7]{Arsenii Galimov}
\author[1]{Ivan Tyukin}
\author[2,8,+]{Rifat Hamoudi}
\author[1,2,+]{Maxim Sharaev}
\affil[1]{Skolkovo Institute of Science and Technology, Moscow, 121205, Russia}
\affil[2]{University of Sharjah, Biomedically Informed Artificial Intelligence Laboratory, Sharjah, 27272, United Arab Emirates}
\affil[3]{Tokai University School of Medicine, Department of Pathology, Isehara, 259-1193, Japan}
\affil[4]{National Research University Higher School of Economics, Moscow, 101000, Russia}
\affil[5]{Lomonosov Moscow State University, Moscow, 119991, Russia}
\affil[6]{Vavilov Institute of General Genetics, Moscow, 119333, Russia}
\affil[7]{Institute of Plant and Animal Ecology Ural Branch of Russian Academy of Sciences, Yekaterinburg, 620144, Russia}
\affil[8]{University of Sharjah, Research Institute for Medical and Health Sciences, Sharjah, 27272, United Arab Emirates}

\affil[*]{Corresponding author: Svetlana Illarionova (s.illarionova@skoltech.ru)}
\affil[+]{these authors contributed equally to this work}

\begin{abstract}

While M2-polarized tumor-associated macrophages (TAMs) have been established as indicators of disease aggressiveness in diffuse large B-cell lymphoma (DLBCL), traditional CD163 immunohistochemistry (IHC) remains resource-intensive. This study aims to investigate the feasibility of using deep learning to quantify TAMs directly from standard hematoxylin and eosin-stained (H\&E) tissue sections. Using a curated dataset of 52 patients with DLBCL, with high-resolution H\&E images and IHC-validated annotations (1,713 TAM instances), five architectures were evaluated: U-Net, Swin-U-Net, Cerberus-U-Net3+, YOLOv11, and HoVer-Net. High CD163 TAM density ($\geq$20.04\%) was associated with significantly reduced overall survival (HR 2.73; 95\% CI: 1.21--6.16; p = 0.012) and progression-free survival (HR 2.88; 95\% CI: 1.23--6.76; p = 0.011). In multivariate Cox proportional hazards analysis adjusting for IPI, molecular subtype (GCB/non-GCB per Hans algorithm), EBV status, and age, CD163 TAM density showed a prognostic trend for overall survival (HR 3.24; 95\% CI: 0.94--11.15; p = 0.062) and progression-free survival (HR 2.41; 95\% CI: 0.80--7.32; p = 0.120). Among the evaluated models, the domain-specific Cerberus-U-Net3+ achieved the highest sensitivity (Recall $0.656 \pm 0.022$), while the Transformer-based Swin-U-Net demonstrated superior segmentation fidelity (Precision $0.694 \pm 0.041$, F1-score $0.633 \pm 0.028$). Additionally, the survival analysis based on the predicted Swin-U-Net CD163 level revealed a 20.8\% cutoff point for patients with high and low CD163 levels near IHC, as well as a downward trend in overall survival among patients with higher predicted CD163 values of 2.63 (95\% CI: 0.85-8.33; p = 0.083). These findings suggest that deep learning architectures using shifted-window self-attention can extract biologically relevant prognostic features from standard histology, and show potential as a candidate surrogate for IHC that is scalable and cost-effective for prognostic assessment of tumor-associated macrophages in diffuse large B-cell lymphoma.


\end{abstract}

\begin{document}
\maketitle
\section*{Introduction}

\begin{figure}[!htbp]
  \centering
  \includegraphics[width=1.\linewidth]{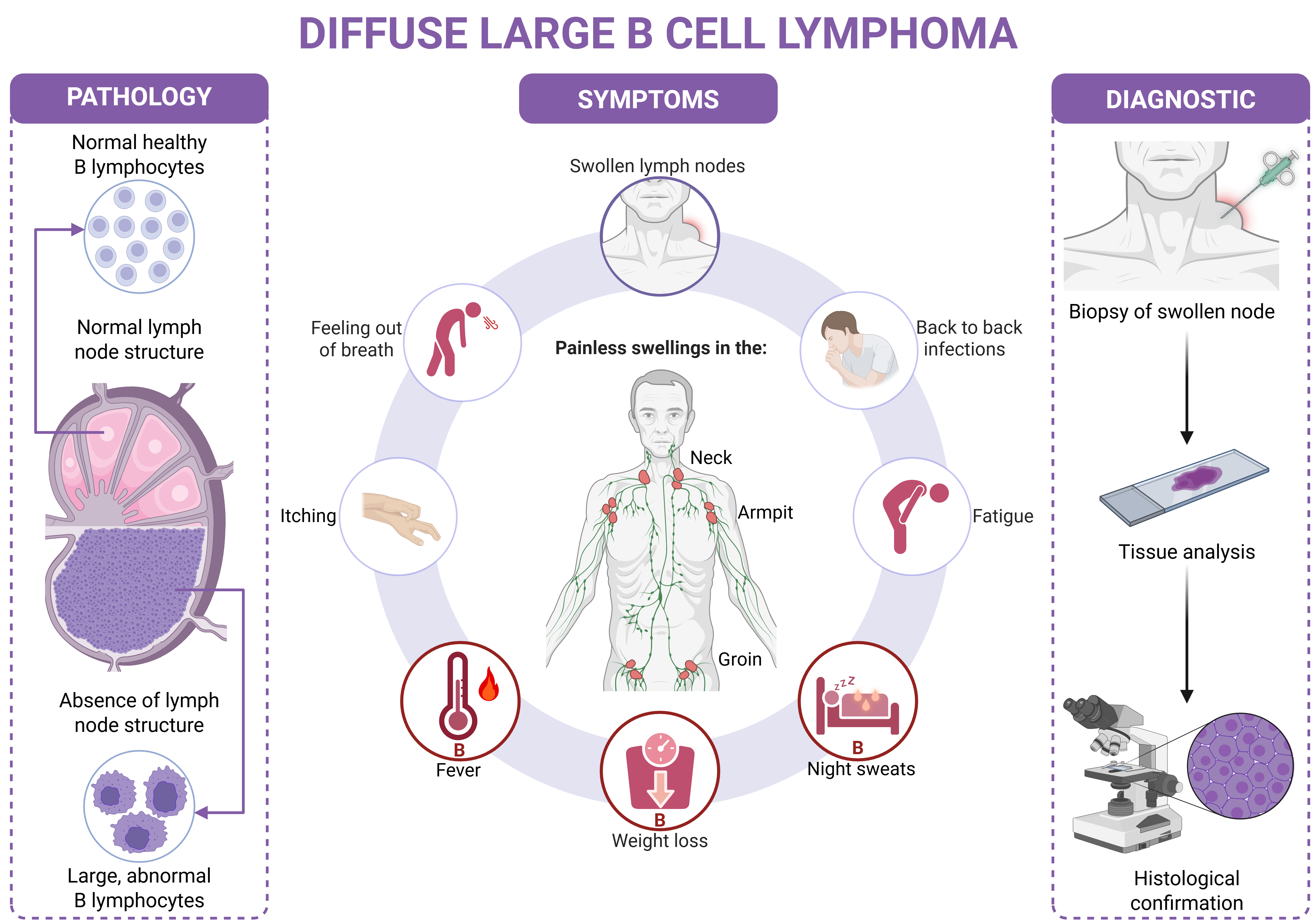}
  \caption{Overview of pathology, clinical manifestations, and diagnostic workflow of diffuse large B-cell lymphoma. The left panel depicts pathological features, illustrating abnormal proliferation of large B lymphocytes within lymphoid tissue. The central panel shows common clinical manifestations, including painless lymphadenopathy, fatigue, recurrent infections, shortness of breath, and systemic “B symptoms” such as fever, night sweats, and weight loss. The right panel outlines the diagnostic workflow, beginning with clinical evaluation and lymph node biopsy, followed by histopathological examination using microscopy. Figure was created using BioRender.com} \label{fig1}
\end{figure}

Diffuse Large B-cell Lymphoma (DLBCL) is an aggressive malignancy that develops from abnormal B-lymphocytes and is characterized by the diffuse proliferation of large neoplastic cells that efface the normal lymph node architecture~\cite{li2018diffuse}.  
Painless lymphadenopathy is the hallmark clinical presentation of DLBCL, frequently involving cervical, axillary, and inguinal sites.
The diagnosis of DLBCL is established on the basis of a pathoanatomic examination of a tumor biopsy~\cite{wu2025comprehensive}. Figure \ref{fig1} summarizes key aspects of DLBCL, including pathology, clinical symptoms, and diagnostic procedures.

It is the most common histological subtype of non-Hodgkin lymphomas, accounting for up to 40\% of cases and estimated 150,000 new cases annually, its risk increases with age and is higher in men than in women~\cite{berhan2025diffuse, wang2023epidemiology}. Despite significant advancements in diagnosis and treatment in recent years, 40–50\% of patients with DLBCL remain incurable, highlighting the urgent need for robust prognostic biomarkers~\cite{xu2024specific}.

The tumor microenvironment, specifically M2-polarized tumor-associated macrophages (TAMs), plays a critical role in immune escape and disease progression. These cells are involved in cancer development, infiltration, dissemination, and immune escape, and strongly associate with unfavorable outcomes in various malignancies, including lymphomas~\cite{funes2018implications}. Clinically, the density of M2 TAMs is identified by the specific marker CD163 ~\cite{lin2019tumor}. High density of CD163 TAMs significantly correlates with poor progression-free and overall survival in DLBCL patients~\cite{lin2023nuclei, cioroianu2019tumor}. Furthermore, high CD163 macrophage infiltration is associated with advanced disease stage and higher International Prognostic Index scores, underscoring its clinical significance~\cite{shen2016m2}. 

While immunohistochemistry (IHC) remains the gold standard for CD163 TAMs quantification, its implementation is hampered by high costs, long turnaround times, and inter-observer variability~\cite{aslani2020comparison, djordjevic2021effects}. In contrast, standard hematoxylin and eosin (H\&E) staining is rapid, inexpensive, and universally available but cannot distinguish between macrophage subtypes, thus failing to provide crucial prognostic information~\cite{oner2025tumor, guo2021correlation, gubanov2026haps}.

The integration of computer vision (CV) algorithms into digital pathology has demonstrated that deep learning architectures - specifically convolutional neural networks (CNNs) and Vision Transformers (ViTs) are capable of extracting complex morphological features from H\&E-stained images~\cite{echle2021deep, illarionova2025hierarchical}. A foundational architecture in this domain is the U-Net, whose symmetric design and skip connections facilitate the preservation of local spatial hierarchies, which are essential for delineating fine cellular boundaries~\cite{ronneberger2015u}. Complementary to semantic segmentation, object detection frameworks such as the YOLO (You Only Look Once) family, including the latest YOLOv11 iterations, enable rapid localization of individual cell instances. These models offer significant advantages in detection tasks, making them potentially applicable to cell detection in H\&E images~\cite{lei2025deep}.
 Graham and co-authors developed HoVer–Net, the first architecture that jointly solves the tasks of instance segmentation and core classification in a single forward pass~\cite{graham2019hover}. This model has become the de facto standard for segmentation in digital pathology.
However, the dense lymphoid environment of DLBCL poses significant challenges for standard architectures. To address these, the Cerberus U-Net, a multi-task learning framework, has been developed to simultaneously perform segmentation and classification, minimizing errors in separating closely apposed macrophages and tumor cells~\cite{graham2023one}. Furthermore, the paradigm shift toward global attention mechanisms has led to the development of the Swin-Tiny U-Net. By leveraging hierarchical Vision Transformers, this model captures long-range dependencies and microenvironmental context, thereby overcoming the inherent locality limitations of standard convolutional layers when analyzing heterogeneous tumor landscape~\cite{cao2022swin}.

While emerging virtual staining platforms like VISTA~\cite{aggarwal2025artificial} can transform H\&E images into surrogate for IHC CD163 macrophages representations, they rely on perfectly co-registered paired slides, introduce computational complexity, and do not inherently provide the quantitative cell-level segmentation required for direct biomarker scoring.

Despite these computational strides, the automated identification of TAMs in H\&E sections remains an unresolved challenge due to several factors. First, macrophages lack a unique or stable morphological phenotype in H\&E staining; their appearance frequently overlaps with other myeloid subsets and reactive lymphocytes. This phenotypic ambiguity, coupled with a critical scarcity of expert-annotated datasets, has hindered the development of reliable diagnostic tools. Existing repositories, such as DLBCL-Morph and LyNSeC, provide valuable data on tumor nuclei and common biomarkers but lack specific labels for the immune-suppressive M2-TAMs landscape~\cite{vrabac2021dlbcl, hussein2023lynsec}.

To address this diagnostic gap, we present a comprehensive comparative evaluation of U-Net, Cerberus U-Net, YOLOv11, Swin-Tiny U-Net, and HoVer-Net for the automated quantification of CD163 TAMs directly from H\&E-stained DLBCL sections. Utilizing a curated dataset of 52 tissue sections with expert-derived ground truth, we investigate whether deep-learned morphological features can serve as a surrogate for CD163 TAMs IHC expression. We test the hypothesis that optimized CV architectures can effectively navigate the phenotypic complexity of the DLBCL microenvironment, providing an accessible and cost-effective tool for prognostic risk stratification. The main contributions of this work are the following:

\begin{itemize}
  \item We collected a meticulously curated dataset of 52 DLBCL patients, comprising high-resolution H\&E and comprehensive clinical outcomes, includes 1,713 manually annotated TAMs instances validated by expert pathologists.
  
  \item We investigate transfer learning from domain-general (ImageNet) and domain-specific (histopathology) pretrained encoders as candidate approaches for H\&E-based M2 TAM quantification, and characterise their relative strengths in precision, recall, and segmentation fidelity, with the aim of serving as a candidate H\&E-based surrogate for CD163 IHC, pending prospective validation.
  
  \item We performed a benchmarking of five deep learning architectures, including Transformer-based Swin-U-Net, the domain-specific Cerberus-U-Net3+, and the landmark segmentation model HoVer-Net, and characterised their relative efficiency in this proof-of-concept clinical task.

  \item We establish the prognostic relevance of CD163 TAM density in this specific cohort as a necessary clinical anchor for the H\&E surrogate, contextualized against the established literature, demonstrating its independent association with adverse overall and progression-free survival after multivariate adjustment.
  
\end{itemize}

\section*{Related work}

Despite promising results achieved over the past five years in detecting and classifying DLBCL using deep learning methods, research in this area remains limited~\cite{luna2025ai}. Existing studies can be categorized into three main groups: detecting DLBCL among other lymphomas, classifying DLBCL subtypes, and integrating classification with prognostic biomarker identification.

Since 2020, several studies have successfully employed CNNs to recognize DLBCL in H\&E-stained slides~\cite{li2020deep}. Gupta and colleagues evaluated multiple deep learning architectures for DLBCL subtype classification, identifying ConvNeXt as the most effective model~\cite{gupta2024classification}. Beyond classification, researchers have proposed segmenting tumor cell nuclei within the complex DLBCL microenvironment. 

The HoVer-Net architecture~\cite{graham2019hover}, the first to jointly solve nuclear instance segmentation and cell-type classification in a single forward pass, using horizontal and vertical distance maps to separate densely packed nuclei. HoVer-Net and its modified version HoLy-Net have been applied to relatively small datasets of fewer than 100 patients~\cite{naji2024holy, tang2025enhanced}. However, HoVer-Net's direct application to DLBCL macrophage segmentation is constrained by its pre-training on solid-tumour datasets: published benchmarks report Dice 0.47--0.58 for the macrophage class, with no differentiation between M1 and M2 subtypes~\cite{zhang2024hovernet}.

Critically, while these works excel at identifying and segmenting neoplastic B-cells, they do not address the specific quantification of tumor-infiltrating immune cells, particularly M2-phenotype TAMs in the DLBCL microenvironment.

On the other hand, significant progress has been made in detecting and quantifying TAMs in solid tumors over the past five years. Most studies focus on TAMs detection using IHC staining for CD163 . Deep learning architectures including DeepLabV3, U-Net, and ResNet18 have achieved high accuracy in automatically identifying CD163 TAMs, with DeepLabV3 demonstrating instance separation efficiency exceeding 89\%~\cite{cancian2021development}.

There are two more areas of research that are: virtual CD163 staining and direct TAM detection in H\&E stained sections. Deep learning algorithms for virtual staining could potentially enable TAM identification from H\&E images. However, a key challenge is obtaining paired IHC and H\&E sections without significant registration errors or artifacts. In 2025, Aggarwal et al. introduced VISTA, a platform that converts H\&E images into virtually stained CD163 images for M2-TAM detection~\cite{aggarwal2025artificial}. The authors reported that VISTA outperforms existing virtual staining methods such as Pyramid Pix2Pix, PSPStain, and ASP, likely because these methods were originally designed for HER2 marker visualization in breast cancer rather than CD163 ~\cite{liu2022bci, chen2024pathological, li2023adaptive}. Currently, virtual staining research focuses predominantly on breast cancer and has not addressed M2 TAM detection in lymphomas. 
Notably, the VISTA model weights are not publicly available, preventing independent reproduction. In addition, VISTA requires precisely co-registered H\&E and IHC slide pairs, whereas our dataset consists of consecutive sections and therefore does not fully satisfy this requirement, precluding its application to M2 macrophage detection in DLBCL in the present study.

No studies to date have combined deep learning approaches for macrophage identification in DLBCL using H\&E-stained sections. One promising architecture for this application is Cerberus~\cite{graham2023one}, developed for simultaneous segmentation and classification of histopathological structures in H\&E images. Cerberus was trained on over 440,000 image patches for tissue classification across multiple independent datasets. According to its developers, Cerberus's capabilities in nuclear segmentation and tissue classification enable accurate determination of tumor-infiltrating immune cells and their spatial distribution. Recent studies demonstrate that Cerberus outperforms standard U-Net baselines in breast and liver cancer tissue analysis~\cite{wang2024multi}. 

The main challenge remains the lack of reliable computer vision systems for identifying and quantifying TAMs in H\&E images. This is mainly due to the fact that macrophages do not have distinct morphological features after staining with hematoxylin, making it difficult to reliably differentiate them from histiocytes, endothelial cells, or other reactive lymphoid elements.
As a result, most current TAM analysis solutions still rely on IHC staining, which reintroduces the original limitations of the method. However, the development of deep learning models for automatic identification, segmentation, and quantification of TAMs on H\&E stained slides represents a promising approach that could address these technological challenges.

The prognostic role of CD163 TAMs in DLBCL has been documented across multiple independent cohorts. Lin et al. reported that CD163 TAM density constitutes an independent prognostic factor in DLBCL, with high infiltration associated with inferior survival in the R-CHOP treatment era~\cite{lin2023nuclei}. Cioroianu et al. demonstrated that elevated CD163 macrophage density correlates with advanced Ann Arbor stage and higher International Prognostic Index (IPI) scores, underscoring the link between macrophage burden and established risk parameters~\cite{cioroianu2019tumor}. Shen et al. further showed that high M2 macrophage infiltration is independently associated with inferior progression-free and overall survival in DLBCL patients receiving R-CHOP~\cite{shen2016m2}. The novelty of the present work lies not in the clinical finding per se, but in demonstrating that a quantitative threshold derived from a standardized digital approach can be reproduced by an H\&E-based deep learning model--a contribution not previously reported for DLBCL.

\section*{Materials}

\subsubsection*{Patient Cohort}

The study cohort comprised 52 patients diagnosed with DLBCL at Tokai University (Japan). The median age was 67 years (IQR: 59–77 years), with a balanced sex distribution (27 males, 25 females). Patients were predominantly treated with R-CHOP (rituximab, cyclophosphamide, doxorubicin, vincristine, and prednisone) or R-CHOP-like regimens. Clinical follow-up documented 26 mortality events - Overall Survival, and 24 Progression-Free Survival events.

\subsubsection*{Dataset Preparation and Annotation}

To establish the ground truth for CD163 TAM quantification, IHC staining was performed using the Bond-Max fully automated IHC and in situ hybridization staining system according to the manufacturer's protocol (Leica Biosystems, Tokyo, Japan). Sections were stained using a polymer detection system (DS9800, Leica Biosystems) with diaminobenzidine (DAB) chromogen and hematoxylin counterstain. The primary antibody, targeting CD163 molecule was Leica anti-CD163 (CD163 -L-CE, Clone 10D6); antigen retrieval was high pH 9 (ER2, Leica Biosystems).

CD163 macrophages in the tumor microenvironment were quantified and expressed as a percentage of total cells. 
Whole-tissue sections were stained with H\&E and digitized using a high-throughput slide scanner (NanoZoomer S360, C13220-01, Hamamatsu Photonics K.K.) at 20× magnification. 
Annotations were manually performed by a trained pathologist under senior supervision using a conservative two-step protocol. CD163 cells were identified on IHC slides using the DAB chromogenic signal as a definitive marker. Morphologically corresponding cells were then located on paired H\&E slides at the same tissue coordinates. Cells were annotated on H\&E only if a morphologically convincing match could be identified. In cases where there was no clear morphological correlate on H\&E, either due to section plane displacement or genuine H\&E ambiguity, the cell was conservatively left unannotated. The spatial offset between consecutive sections was typically subcellular at 20× magnification, and expert annotators used IHC as a spatial reference rather than a pixel-exact overlay. The IHC served solely as a spatial reference guide, and no automated IHC annotation was used to generate ground-truth masks.

The resulting dataset comprised 52 representative Regions of Interest (ROIs), each measuring ($2560 \times 1456$ pixels), and contained 1,713 individually annotated macrophages. These annotations were converted into binary masks.
The CD163 percentage was calculated as CD163 macrophage pixel area to total nucleated cell area, which were computed via the StarDist QuPath extension.
The training dataset was constructed using the following approach: High-resolution images $2560 \times 1456$ pixels were subdivided into non-overlapping $256 \times 256$-pixel patches with corresponding binary masks. 
To prevent edge truncation artifacts, each annotated TAM was centered within its patch using geometric transformations, avoiding incomplete macrophage representations during model training. To maintain a balanced representation and enable robust negative sampling, 1,145 background patches were randomly extracted from regions devoid of CD163 TAMs infiltration. The final dataset comprised 1,142 TAM-positive and 1,145 background patches.

\begin{figure}[!htbp]
  \centering
  \includegraphics[width=1.\linewidth]{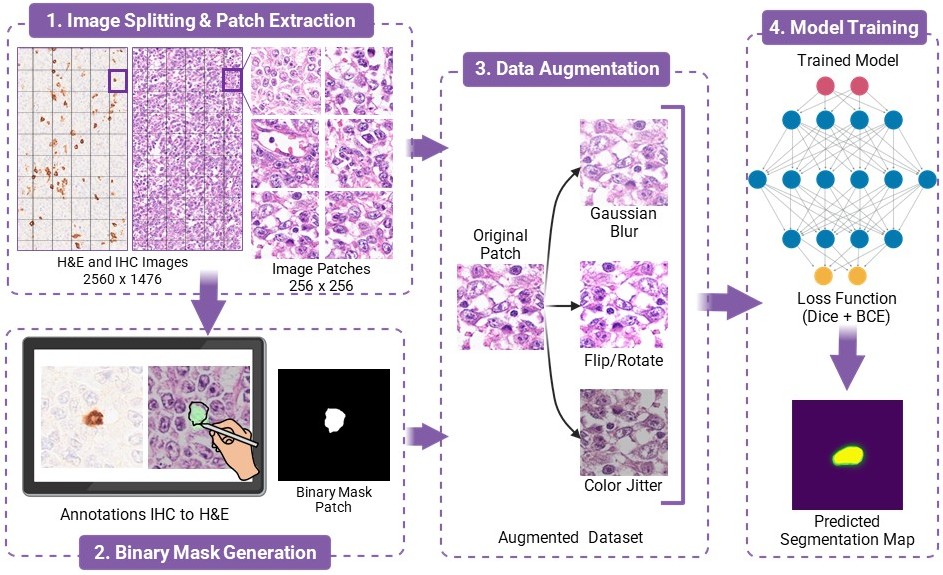}
  \caption{Workflow for histopathological image processing, annotation, and model training.
  The workflow illustrates the sequential steps used for image-based macrophage analysis. ROIs (2560 × 1456 pixels) from paired H\&E and IHC CD163 slides are subdivided into smaller image patches with macrophages. Binary mask generation is performed via manual macrophages annotations on H\&E images using the corresponding IHC CD163 signal as a definitive spatial reference. Data augmentation is applied to annotated patches to increase dataset variability, including transformations such as rotation and intensity variation. The augmented image set is used for model training, where a neural network is trained to generate prediction outputs, visualized as probability or activation maps.
  Figure was created using BioRender.com}\label{fig2}
\end{figure}

The workflow of macrophage analysis is schematically shown in Figure \ref{fig2}.

\subsubsection*{Data Augmentation}

For each macrophage-containing patch, ten additional augmented images were generated using a transformation pipeline implemented with the Albumentations~\cite{buslaev2020albumentations}. For each positive patch, ten augmented variants were generated. The pipeline included:

\begin{itemize}
  \item \textbf{Geometric transformations:} Horizontal and vertical flips (p = 0.5) to ensure rotation invariance.
  \item \textbf{Chromatic augmentation (Color Jittering):} Random adjustments to brightness (0.01–0.3), contrast (0.01–0.3), saturation (0.01–0.4), and hue (0.01–0.1) to account for staining variability (p = 0.5).
  \item \textbf{Blurring and Noise:} Gaussian blur with a 3×3 kernel (p = 0.5) to simulate variations in focus and image degradation.
\end{itemize}

All transformations were applied stochastically to ensure a diverse augmented dataset while preserving the semantic integrity of the original images~\cite{nesteruk2024image}. 

\section*{Methods}

The primary objective of this study is to bridge the diagnostic gap between the clinical necessity for CD163 TAM quantification and the practical limitations of IHC in DLBCL management. We framed this as an extraction of a sub-visual macrophage pattern directly from standard H\&E-stained images. To identify the optimal algorithmic approach for this task, we conducted a comprehensive comparative evaluation of four distinct deep learning paradigms:

\begin{enumerate}

 \item We selected U-Net as our primary baseline benchmark because its often used for medical image segmentation, ensuring the preservation of high-resolution spatial features necessary for identifying cellular boundaries~\cite{ronneberger2015u}.

 \item We selected a Swin-U-Net architecture (an ImageNet-pretrained Swin-Tiny encoder paired with a lightweight U-Net-style decoder) for TAM segmentation in H\&E slides for several reasons. First, Swin Transformer provides a hierarchical multi-scale representation with shifted-window self-attention, which retains computational efficiency while enabling cross-window interactions and strong performance on dense prediction tasks, making it a suitable backbone when both fine-grained boundaries and broader tissue context matter~\cite{liu2021swin}. Swin-U-Net specifically demonstrates that combining Swin-style context modeling with a U-Net-like decoding pathway can improve medical segmentation accuracy by learning both local and long-range semantic interactions, which might be critical for differentiating pleomorphic TAMs from the dense background of tumor cells~\cite{cao2022swin}.

 \item To address the inherent scarcity of large-scale annotated datasets in rare lymphoma subtypes, we evaluated the Cerberus-U-Net framework. This model utilizes a multi-task, domain-specific backbone pre-trained on diverse histopathological datasets, hypothesized to provide more robust and generalizable feature representations than standard ImageNet-based pre-training~\cite{graham2023one}.

 \item We included the YOLOv11-seg architecture~\cite{khanam2024yolov11} evaluate the efficacy of state-of-the-art instance segmentation. Unlike semantic approaches, YOLO-based models treat each macrophage as a discrete object, potentially offering superior performance in high-throughput clinical workflows where rapid cell counting and localization are paramount~\cite{lei2025deep}.
 
 \item To benchmark against a representative domain-specific cell segmentation algorithm, we additionally evaluated HoVer-Net~\cite{graham2019hover}, a de facto standard for nuclear instance segmentation in digital pathology, initialized from PanNuke-pretrained weights. This allows us to contextualise the performance of our purpose-trained models relative to an off-the-shelf domain-specific baseline with established performance in histopathological segmentation tasks.

\end{enumerate}

\subsubsection*{Deep Learning Architectures and Training Protocol}

To assess the quality of the experiments, we used 5-fold cross-validation ($k=5$) with folds defined at the parent-image level, where each parent-image corresponds to an individual patient. First, all original (non-augmented) patches were grouped by their parent image, and parent images were assigned to one of five folds such that all patches from the same parent image remained within a single fold. In each run, four folds (41 parent images) were used for training and the remaining fold (11 parent images) was used for validation. Data augmentation was applied only to training patches within each fold (as described in the Data Augmentation section), while validation was performed on the corresponding non-augmented patches. The model was trained for 30 epochs per fold. This strict parent-image partitioning prevents overly optimistic estimates that can arise when highly similar patches originating from the same image inadvertently appear in both training and validation splits.

For optimization, we use AdamW, which decouples weight decay from the gradient-based update and is widely adopted for transformer-based training due to improved regularization behavior compared to coupling weight decay with adaptive updates~\cite{loshchilov2017decoupled}. 

The objective function was a hybrid loss (1:1 ratio) combining BCEWithLogits Loss and Dice Loss. Inference thresholded sigmoid outputs at 0.5. This dual approach provides stable pixel-wise gradients from cross-entropy-style supervision and directly optimize overlap under class imbalance, a common and empirically effective practice in segmentation settings~\cite{yeung2022unified,hosseini2024topk}.

\subsubsection*{Baseline Segmentation: U-Net}

We implemented a standard U-Net architecture using the segmentation\_models\_pytorch library~\cite{ronneberger2015u}. To optimize the balance between depth and parameter efficiency, an EfficientNet-Lite backbone (approximately 3 million parameters) was utilized, initialized with ImageNet weights. The lightweight nature of this encoder was selected to minimize the risk of overfitting on our specialized DLBCL dataset while maintaining the multi-scale feature fusion capabilities provided by the U-Net’s symmetric skip connections.

\subsubsection*{Transformer-based Segmentation: Swin-U-Net}

Our primary model coupled a swin\_tiny\_patch4\_window7\_224 ImageNet-pretrained encoder with a lightweight U-Net-style decoder ("Swin-U-Net")~\cite{cao2022swin}. ImageNet mean/standard deviation normalization was used for the encoder input. Swin-U-Net specifically demonstrates that combining Swin-style context modeling with a U-Net-like decoding pathway can improve medical segmentation accuracy by learning both local and long-range semantic interactions, which might be critical for differentiating pleomorphic TAMs from the dense background of tumor cells~\cite{cao2022swin}.

\subsubsection*{Domain-Specific Transfer Learning: Cerberus}

Recognizing the limitations of general ImageNet pre-training for histopathological tasks, we initialized the encoder using weights derived from Cerberus, which were pre-trained on diverse histopathological datasets. Specifically, a ResNet-34 encoder pre-trained within the Cerberus multi-task learning paradigm was employed as a feature extraction backbone~\cite{graham2023one}. To leverage the generalized representations of tissue morphology learned during multi-domain pre-training, the encoder weights (21.2 million parameters) were frozen. We then evaluated three decoder variants—standard U-Net, U-Net++, and U-Net3+ to identify the optimal architecture for recovering spatial resolution from domain-specific features. 

\subsubsection*{Instance Segmentation: YOLOv11}

We evaluated the YOLOv11-seg family (Nano, Small, Medium, and Extra-Large) to assess performance through the lens of instance-level localization~\cite{lei2025deep}. Training employed the stochastic gradient descent (SGD) optimizer with initial learning rate 0.01, momentum 0.937, and regularization weight decay 0.0005. Learning rate was controlled using cosine scheduling with minimum rate 0.01 and 3-epoch warmup phase.

\subsubsection*{Domain-Specific Cell Segmentation: HoVer-Net}

We additionally evaluated HoVer-Net~\cite{graham2019hover}, the original architecture comprises a pre-activated ResNet-50 backbone and three parallel decoder branches: a binary nuclear pixel (NP) branch, a horizontal/vertical distance map (HV) branch, and a nuclear type classification (NC) branch. HoVer-Net has become the de facto standard for nuclear segmentation in digital pathology and has been applied to lymphoid tissue in several recent studies~\cite{naji2024holy, tang2025enhanced}. In the present implementation, ResNet-50 was used as the encoder and initialized with weights pre-trained on PanNuke~\cite{gamper2020pannuke}. The total number of trainable parameters was 43.9M; with the encoder frozen, 20.4M parameters were optimized.

\subsubsection*{Evaluation Metrics}

Model performance was assessed using standard semantic segmentation metrics. All metrics are computed at the pixel level, comparing predicted binary segmentation masks against ground-truth binary masks within each $256 \times 256$ patch; they are not instance-level metrics. Intersection over union (IoU) quantifies the ratio of intersection to union of predicted and true positive regions. Precision represents the proportion of true positives among all positive predictions, while Recall measures the proportion of true positives among all actual positive instances. F1 score provides the harmonic mean of Precision and Recall. For models evaluated with k-fold cross-validation, metrics were computed by aggregating true positives, false positives, and false negatives across all validation samples within each fold, with mean and standard deviation reported across folds.

\begin{align}
\text{IoU} &= \frac{|A \cap B|}{|A \cup B|} = \frac{TP}{TP + FP + FN} \\
\text{Precision} &= \frac{TP}{TP + FP} \\
\text{Recall} &= \frac{TP}{TP + FN} \\
\text{F1} &= \frac{2 \times \text{Precision} \times \text{Recall}}{\text{Precision} + \text{Recall}} = \frac{2 \times TP}{2 \times TP + FP + FN}
\end{align}

Where: $TP$ - True Positive, $FP$ - False Positive, $FN$ - False Negative, $A$ - Ground Truth mask, $B$ - Predicted mask.

\subsection*{Statistical Analysis}

Clinical outcome analysis focused on two primary endpoints: Overall Survival, defined as the time from diagnosis to death from any cause, and Progression-Free Survival, defined as the time to disease progression, relapse, or death. To categorize patients into "High" and "Low" macrophage infiltration groups, we determined the optimal threshold for TAM density using maximally selected log-rank statistics (the maxstat algorithm). This approach minimizes subjective bias in cutoff selection by identifying the point of maximum divergence in survival probability.

The same maxstat algorithm was applied independently to the predicted CD163 area-fraction values derived from the best deep learning model to determine model-specific optimal cutpoints. To assess the clinical validity the IHC-derived threshold of CD163 was applied directly to the best model's continuous output to classify patients as high/low, and concordance with IHC-derived binary classification was measured using Cohen's $\kappa$ coefficient. Linear regression (model output $\sim$ IHC reference) was also performed to evaluate the continuous-level agreement.

Survival curves were estimated using the Kaplan-Meier method, and differences between groups were compared using the log-rank test~\cite{xie2005adjusted}. To evaluate the independent prognostic value of CD163 TAM scores, we employed Cox proportional hazards regression models. We first performed univariate analyses to estimate Hazard Ratios (HR) and their corresponding 95\% Confidence Intervals (CI) for TAM density and a p-value < 0.05 was considered statistically significant. 

To establish independence from established clinical risk factors, multivariate Cox proportional hazards analysis was additionally performed. The following covariates were included: CD163 TAM density (high vs\ low at the cutoff), International Prognostic Index (IPI; low 0--2 vs high 3--5), molecular subtype by Hans algorithm (GCB vs non-GCB), EBV status by EBER in-situ hybridisation (negative vs positive), and age (continuous). Ann Arbor stage and serum LDH levels were not systematically available for all 52 patients in this retrospective cohort and were therefore not included as independent covariates; however, as IPI is a validated composite index that structurally incorporates both Ann Arbor stage III--IV and elevated LDH as two of its five components, its inclusion provides a principled surrogate adjustment for both variables simultaneously.

Statistical analyses were performed using R software, packages (survival, survminer, maxstat)~\cite{chambers2008software}.

\section*{Results}

\subsection*{Survival analysis stratified by CD163 expression}

\begin{figure}[!htbp]
  \centering
  \includegraphics[width=\linewidth]{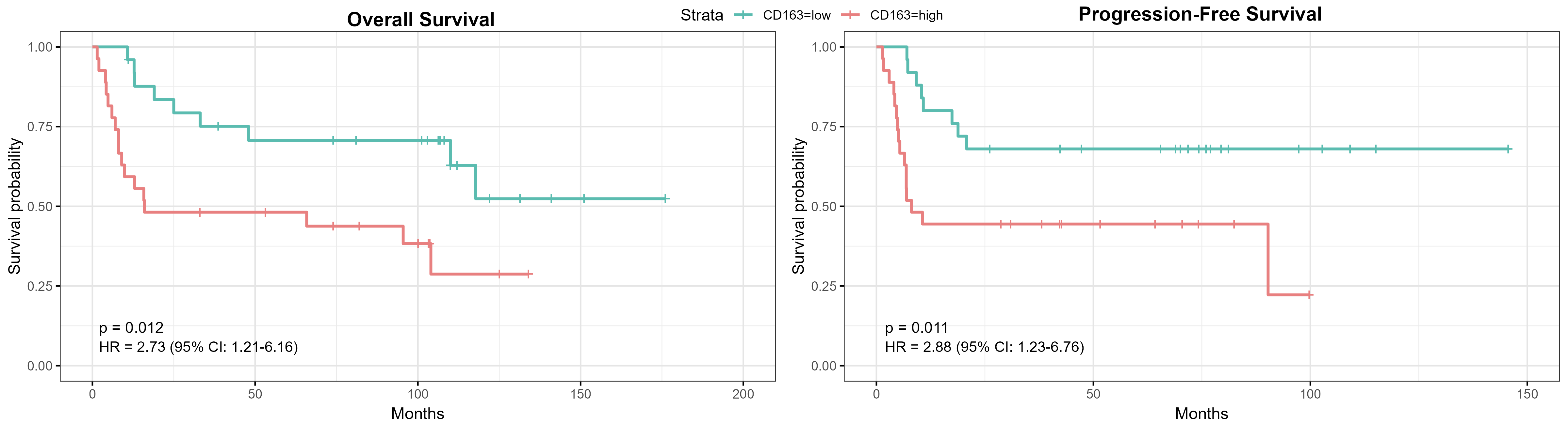}
  \caption{Association of CD163 expression with overall and progression-free survival. Kaplan–Meier curves showing overall survival and progression-free survival in patients stratified according to CD163 expression levels. Patients were classified into CD163 -high (red) and CD163 -low (blue) groups. Survival probability is shown on the y-axis, and time in months is shown on the x-axis.} \label{fig3}
\end{figure}

Using the maximally selected log-rank method, an optimal CD163 cutoff value of 20.04\% was determined, enabling stratification of 52 patients with DLBCL into two groups: 24 patients had values below the defined threshold, and 28 had values equal to or above this cutoff. Kaplan–Meier survival analysis of the 52 patients with DLBCL demonstrated significant differences in both overall survival and progression-free survival between patients with high and low CD163 expression (Figure \ref{fig3}). Patients in the high-CD163 group exhibited reduced overall survival compared with those in the low-CD163 group, with a HR of 2.73 (95\% CI: 1.21–6.16; p = 0.012 by log-rank test). Similarly, progression-free survival was significantly shorter in the high-CD163 group, corresponding to a HR of 2.88 (95\% CI: 1.23–6.76; p = 0.011 by log-rank). These findings indicate an association between elevated CD163 expression and poorer survival outcomes.

The results of multivariate Cox proportional hazards analysis (Tab \ref{cox1}). Covariates: CD163 TAM density (high vs low), IPI (high 3--5 vs low 0--2), Hans molecular subtype (non-GCB vs GCB), EBER status (positive vs negative), and age (continuous). Ann Arbor stage and LDH were not individually available; IPI was used as a validated composite surrogate. Proportional hazards assumption confirmed by Schoenfeld residual analysis (global test: OS p = 0.94; PFS p = 0.33).

\begin{table}[H]
\centering
\caption{Multivariate Cox proportional hazards analysis results.}
\label{cox1}
\setlength{\tabcolsep}{5pt}
\begin{tabular}{lcccc}
\toprule
& \multicolumn{2}{c}{Overall Survival} & \multicolumn{2}{c}{Progression-Free Survival} \\
\cmidrule(lr){2-3}\cmidrule(lr){4-5}
Variable & HR (95\% CI) & p & HR (95\% CI) & p \\
\midrule
CD163 High vs.\ Low & \textbf{3.24 (0.94--11.15)} & \textbf{0.062} & \textbf{2.41 (0.80--7.32)} & \textbf{0.120} \\
IPI High (3--5) vs.\ Low & 1.85 (0.75--4.54) & 0.181 & 1.54 (0.61--3.91) & 0.364 \\
Non-GCB vs.\ GCB (Hans) & 0.57 (0.19--1.72) & 0.315 & 0.63 (0.20--1.97) & 0.432 \\
EBER Positive vs.\ Negative & 1.41 (0.38--5.27) & 0.607 & 1.98 (0.61--6.38) & 0.254 \\
Age (per 1 year) & 1.03 (0.99--1.06) & 0.112 & 1.01 (0.98--1.04) & 0.515 \\
\bottomrule
\end{tabular}
\end{table}

\subsection*{Baseline U-Net}

The quantitative performance of the baseline U-Net under 5-fold cross-validation is summarized in Table \ref{tab1}. The model achieved a mean F1 of ($0.587 \pm 0.038$), demonstrating moderate segmentation fidelity across the validation folds. Notably, the U-Net architecture exhibited higher Recall ($0.635 \pm 0.055$) relative to its Precision ($0.547 \pm 0.034$). While this ensures higher sensitivity in identifying potential macrophage regions, it resulted in a higher rate of false-positive pixel classifications in dense lymphoid areas.

\begin{table}[H]
\centering
\setlength{\tabcolsep}{6pt}
\begin{tabular}{lcccccc}
\toprule
Metric & Mean $\pm$ SD & Fold 1 & Fold 2 & Fold 3 & Fold 4 & Fold 5 \\
\midrule
IoU    & 0.416 $\pm$ 0.038 & 0.422 & 0.380 & 0.436 & 0.467 & 0.377 \\
Precision & 0.547 $\pm$ 0.034 & 0.547 & 0.546 & 0.546 & 0.598 & 0.500 \\
Recall   & 0.635 $\pm$ 0.055 & 0.649 & 0.555 & 0.685 & 0.680 & 0.606 \\
F1     & 0.587 $\pm$ 0.038 & 0.593 & 0.551 & 0.608 & 0.636 & 0.548 \\
\bottomrule
\end{tabular}
\caption{Baseline U-Net performance under 5-fold TAMs segmentation cross-validation}
\label{tab1}
\end{table}

\subsection*{Swin-U-Net}

Swin-U-Net demonstrates higher rates of macrophage segmentation Table \ref{tab2} compared to the baseline U-Net Table \ref{tab1}. Although the Recall of Swin-U-Net was lower than the baseline U-Net ($0.583 \pm 0.034$ vs $0.635 \pm 0.055$), the higher overall spatial overlap in the Swin-U-Net's IoU ($0.464 \pm 0.030$) and F1 ($0.633 \pm 0.028$) scores indicated a more reliable and anatomically accurate mapping of TAMs in DLBCL. The reduced standard deviation across folds suggests that the transformer-based approach is less sensitive to staining variations inherent in the DLBCL dataset.

\begin{table}[H]
\centering
\setlength{\tabcolsep}{6pt}
\begin{tabular}{lcccccc}
\toprule
Metric & Mean $\pm$ SD & Fold 1 & Fold 2 & Fold 3 & Fold 4 & Fold 5 \\
\midrule
IoU    & 0.464 $\pm$ 0.030 & 0.459 & 0.421 & 0.489 & 0.504 & 0.443 \\
Precision & 0.694 $\pm$ 0.041 & 0.642 & 0.677 & 0.751 & 0.732 & 0.670 \\
Recall   & 0.583 $\pm$ 0.034 & 0.617 & 0.527 & 0.584 & 0.619 & 0.567 \\
F1     & 0.633 $\pm$ 0.028 & 0.629 & 0.593 & 0.657 & 0.671 & 0.614 \\
\bottomrule
\end{tabular}
\caption{Swin-U-Net performance under 5-fold TAMs segmentation cross-validation}
\label{tab2}
\end{table}

\subsection*{YOLOv11 Model Family}

Table \ref{tab3} shows segmentation performance indicators for four YOLOv11-based segmentation models. The most compact variant, YOLO11n-seg, achieved the highest F1 score ($0.590 \pm 0.033$) and Recall ($0.641 \pm 0.033$), suggesting that for macrophage segmentation in H\&E sections, increased parameter depth does not necessarily translate to improved feature extraction. Larger variants like YOLO11s-seg and YOLO11m-seg achieved higher Precision ($0.736$ and $0.729$, respectively) but at the cost of significantly lower Recall. Notably, increasing the model complexity from nano to extra-large variants did not result in a substantial improvement in performance, and the largest model (YOLOv11x-seg) did not demonstrate any advantage over more compact versions. Therefore, YOLO11n-seg achieved the highest F1 score, and this option, YOLO11n-seg, was selected for the final model comparison.

\begin{table}[H]
\centering
\setlength{\tabcolsep}{5pt}
\begin{tabular}{lccccccc}
\toprule
Metric & Variant & Mean $\pm$ SD & Fold 1 & Fold 2 & Fold 3 & Fold 4 & Fold 5 \\
\midrule
 & n-seg & $0.466 \pm 0.037$ & 0.412 & 0.454 & 0.491 & 0.508 & 0.466\\
IoU& s-seg & $0.463 \pm 0.034$ & 0.460 & 0.410 & 0.493 & 0.492 & 0.460\\
 & m-seg & $\mathbf{0.469 \pm 0.035}$ & 0.465 & 0.413 & 0.477 & 0.506 & 0.485\\
 & x-seg & $0.466 \pm 0.028$ & 0.461 & 0.421 & 0.487 & 0.492 & 0.467\\
\midrule
 & n-seg & $0.665 \pm 0.089$ & 0.515 & 0.659 & 0.705 & 0.743 & 0.703\\
Precision& s-seg & $\mathbf{0.736 \pm 0.037}$ & 0.694 & 0.744 & 0.723 & 0.793 & 0.726\\
 & m-seg & $0.729 \pm 0.024$ & 0.730 & 0.724 & 0.726 & 0.766 & 0.698\\
 & x-seg & $0.706 \pm 0.058$ & 0.738 & 0.609 & 0.695 & 0.742 & 0.745\\
\midrule
 & n-seg & $\mathbf{0.641 \pm 0.033}$ & 0.692 & 0.622 & 0.647 & 0.639 & 0.605\\
Recall& s-seg & $0.588 \pm 0.048$ & 0.610 & 0.512 & 0.641 & 0.593 & 0.583\\
 & m-seg & $0.601 \pm 0.042$ & 0.597 & 0.530 & 0.618 & 0.625 & 0.636\\
 & x-seg & $0.611 \pm 0.030$ & 0.593 & 0.615 & 0.658 & 0.613 & 0.578\\
\midrule
 & n-seg & $\mathbf{0.590 \pm 0.033}$ & 0.546 & 0.579 & 0.615 & 0.630 & 0.579\\
F1& s-seg & $0.577 \pm 0.040$ & 0.579 & 0.515 & 0.617 & 0.605 & 0.569\\
 & m-seg & $0.586 \pm 0.039$ & 0.584 & 0.520 & 0.598 & 0.625 & 0.602\\
 & x-seg & $0.584 \pm 0.030$ & 0.577 & 0.542 & 0.615 & 0.612 & 0.575\\
\bottomrule
\end{tabular}
\caption{YOLOv11-seg family results under 5-fold TAMs segmentation cross-validation}
\label{tab3}
\end{table}

\subsection*{Transfer Learning with Cerberus}

Based on the presented results Table \ref{tab4} containing segmentation performance metrics for three U-Net–based architectures. The architectures U-Net3+ was selected due to its strong overall segmentation performance. U-Net3+ achieved consistently high quantitative metrics, including competitive IoU values and the highest Recall among the evaluated U-Net–based architectures, indicating reliable detection of target regions.

\begin{table}[H]
\centering
\setlength{\tabcolsep}{5pt}
\begin{tabular}{lccccccc}
\toprule
Metric & Variant & Mean $\pm$ SD & Fold 1 & Fold 2 & Fold 3 & Fold 4 & Fold 5 \\
\midrule
 & U-Net & $0.445 \pm 0.020$ & 0.451 & 0.410 & 0.466 & 0.474 & 0.444\\
IoU & U-Net++ & $\mathbf{0.479 \pm 0.018}$ & 0.461 & 0.451 & 0.491 & 0.493 & 0.458\\
 & U-Net3+ & $0.471 \pm 0.021$ & 0.449 & 0.427 & 0.469 & 0.490 & 0.449\\
\midrule
 & U-Net & $\mathbf{0.712 \pm 0.015}$ & 0.634 & 0.628 & 0.647 & 0.676 & 0.628\\
Precision & U-Net++ & $0.702 \pm 0.014$ & 0.670 & 0.686 & 0.716 & 0.705 & 0.684\\
 & U-Net3+ & $0.705 \pm 0.014$ & 0.630 & 0.620 & 0.658 & 0.673 & 0.626\\
\midrule
 & U-Net & $0.574 \pm 0.022$ & 0.598 & 0.514 & 0.631 & 0.606 & 0.581\\
Recall & U-Net++ & $0.627 \pm 0.020$ & 0.630 & 0.573 & 0.633 & 0.652 & 0.615\\
 & U-Net3+ & $\mathbf{0.656 \pm 0.022}$ & 0.600 & 0.566 & 0.629 & 0.638 & 0.595\\
\midrule
 & U-Net & $0.569 \pm 0.018$ & 0.575 & 0.523 & 0.593 & 0.598 & 0.562\\
F1 & U-Net++ & $0.602 \pm 0.016$ & 0.586 & 0.548 & 0.615 & 0.620 & 0.579\\
 & U-Net3+ & $\mathbf{0.607 \pm 0.018}$ & 0.576 & 0.549 & 0.597 & 0.619 & 0.568\\
\bottomrule
\end{tabular}
\caption{Comparative segmentation performance of different U-net architectures across Cerberus experiments}
\label{tab4}
\end{table}

\subsection*{HoVer-Net} 
The quantitative performance of the HoVer-Net implementation under 5-fold cross-validation is summarized in Table~\ref{tab5}. With a pre-trained PanNuke encoder and a stripped NP-branch decoder, HoVer-Net achieved a mean F1 of $0. 613 \pm 0.018$ and IoU of $0.442 \pm 0.019$. Precision ($0.598 \pm 0.032$) was moderately higher than that of the baseline U-Net ($0.547 \pm 0.034$), reflecting the benefit of domain-specific pre-training in reducing false-positive pixel classifications. However, the model's Recall ($0.636 \pm 0.028$) was broadly comparable to U-Net, and its F1 score remained below those of the Swin-U-Net and Cerberus-U-Net3+. 

\begin{table}[H]
\centering
\setlength{\tabcolsep}{6pt}
\begin{tabular}{lcccccc}
\toprule
Metric & Mean $\pm$ SD & Fold 1 & Fold 2 & Fold 3 & Fold 4 & Fold 5 \\
\midrule
IoU    & 0.442 $\pm$ 0.019	& 0.438 & 0.430 & 0.453 & 0.469 & 0.420 \\
Precision & 0.598 $\pm$ 0.032 & 0.590 & 0.573 & 0.602 & 0.650 & 0.573 \\
Recall   & 0.636 $\pm$ 0.028 & 0.655 & 0.611 & 0.657 & 0.659 & 0.601 \\
F1     & 0.613 $\pm$ 0.018 &	0.609 &	0.601 &	0.623 &	0.638 &	0.592\\
\bottomrule
\end{tabular}
\caption{HoVer-Net performance under 5-fold TAMs segmentation cross-validation}
\label{tab5}
\end{table}

 \subsection*{Comparative Analysis of Segmentation Models}
 
We evaluated the performance of five distinct segmentation approaches: a baseline U-Net, a Swin-Transformer-based model (Swin-U-Net), an instance segmentation model (YOLOv11), a transfer learning approach using a foundation histopathology encoder (ResNet34) with a U-Net decoder (Cerberus-U-Net3+), and the domain-specific cell segmentation algorithm HoVer-Net. These are shown in Table~\ref{tab6}. The baseline U-Net showed the lowest F1 score of 0.587, indicating limited overall segmentation accuracy. YOLOv11n-seg provided a small improvement over the baseline, reaching an F1 score of 0.590 and Recall of 0.641. HoVer-Net, initialized from PanNuke weights, Precision of 0.598, while remaining below the specialized architectures overall. The Cerberus+U-Net3+ transfer learning approach demonstrated the highest Recall of 0.656 and IoU of 0.471, indicating improved sensitivity and spatial overlap. The Swin-U-Net achieved the highest F1 score of 0.633 together with the highest Precision of 0.694, reflecting more accurate boundary delineation and fewer false positives.

\begin{table}[ht]
  \centering
  \caption{Comparative performance metrics of all segmentation models. Bold indicates the best performance for each metric. All metrics are reported as mean $\pm$ SD across 5-fold cross-validation.}
  \label{tab6}
  \begin{tabular}{lccccc}
    \toprule
    Metric & Baseline U-Net & HoVer-Net & YOLO11n-seg & Cerberus U-Net3+ & Swin-U-Net \\
    \midrule
    IoU       & $0.416 \pm 0.038$ & $0.442 \pm 0.019$ & $0.466 \pm 0.037$ & $\mathbf{0.471 \pm 0.021}$ & $0.464 \pm 0.030$ \\
    Precision & $0.547 \pm 0.034$ & $0.598 \pm 0.032$ & $0.665 \pm 0.088$ & $0.628 \pm 0.025$ & $\mathbf{0.694 \pm 0.041}$ \\
    Recall    & $0.635 \pm 0.054$ & $0.636 \pm 0.028$ & $0.641 \pm 0.033$ & $\mathbf{0.656 \pm 0.022}$ & $0.583 \pm 0.034$ \\
    F1        & $0.587 \pm 0.038$ & $0.613 \pm 0.018$ & $0.590 \pm 0.033$ & $0.607 \pm 0.018$ & $\mathbf{0.633 \pm 0.028}$ \\
    \bottomrule
  \end{tabular}
\end{table}

\begin{figure}[!htbp]
  \centering
  \includegraphics[width=0.8\linewidth]{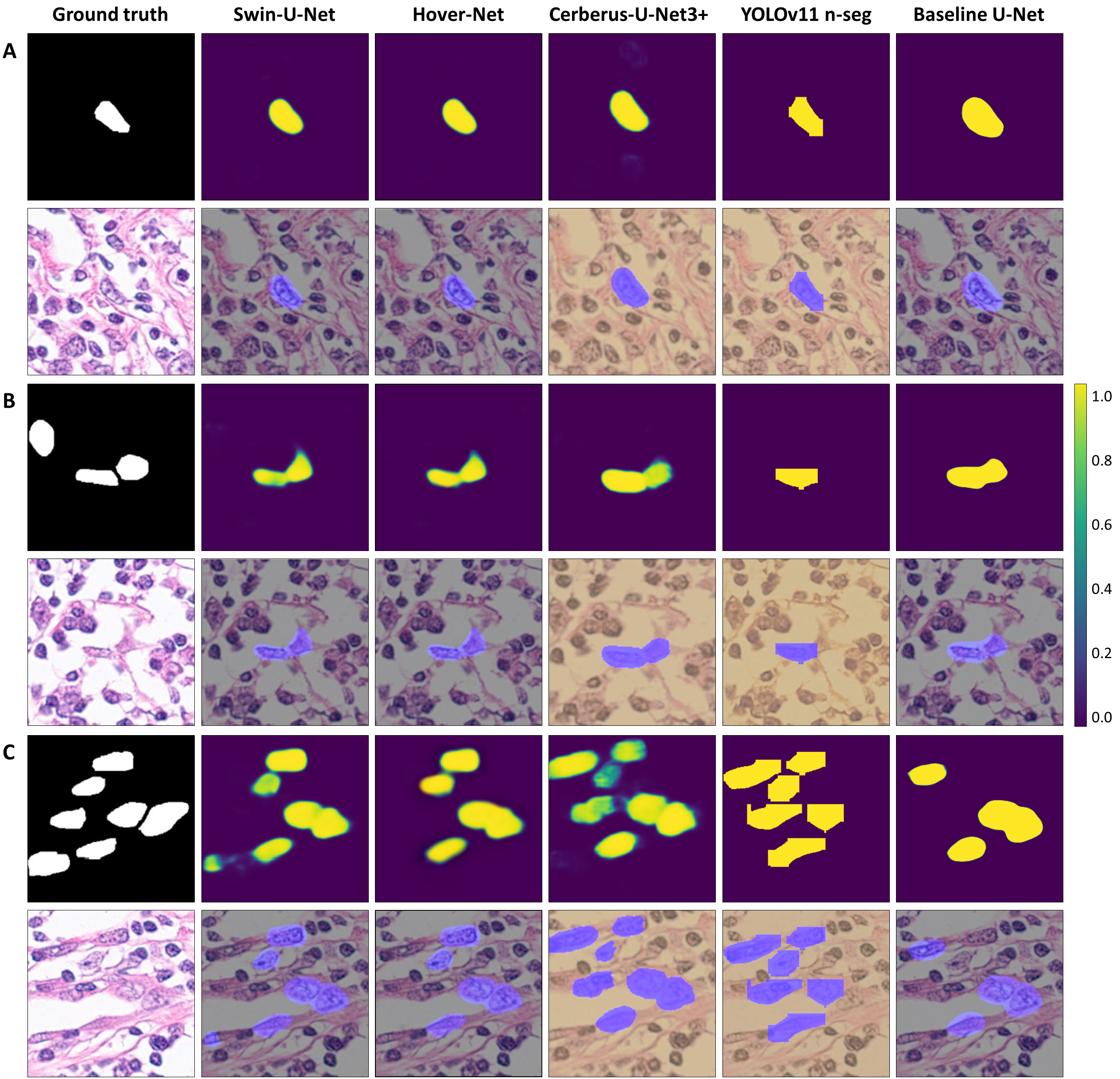}
  \caption{Visual comparison of segmentation predictions for representative histopathological image patches. \\
  For each case, the first column displays the ground truth binary mask. Predicted probability maps are visualized using a color scale ranging from 0 to 1, as indicated by the color bar. For each model, the upper row shows the prediction output, and the lower row shows the predicted segmentation overlaid on the original histopathological image.} \label{fig4}
\end{figure}

Figure \ref{fig4} shows a qualitative comparison of the results of macrophage segmentation obtained using four deep learning models in three representative areas of histopathological images. All models perform well in segmenting single macrophages, quite clearly marking the contours of the cell, Figure \ref{fig4} (A). But in more complex cases, if there are several macrophages on a patch, especially if they are located in a tight cluster, Swin-U-Net detects them more accurately. Although it misses some, this is compensated by the fact that it produces very few false positives, unlike other models Figure \ref{fig4} (C). YOLOv11 n-seg it produced more discrete and rigid segmentation results, but demonstrated less accurate shape matching in complex scenarios.

Clinical outcome analysis focused on two primary endpoints: Overall Survival, defined as the time from diagnosis to death from any cause, and Progression-Free Survival, defined as the time to disease progression, relapse, or death. To categorize patients into "High" and "Low" macrophage infiltration groups, we determined the optimal threshold for TAM density using the maxstat algorithm applied to the Swin-U-Net predicted CD163 area-fraction values across the 52 patients, employing an identical statistical procedure to that used for the IHC-derived threshold.

\subsection*{The prognostic value of the Swin-U-Net predicted CD163 quantification}
To evaluate the feasibility of a digital surrogate based on H\&E, survival analysis was repeated using Swin-U-Net predictions of CD163 . The optimal cutpoint derived from the model (20.8\%) was close to the threshold derived from IHC (20.04\%). 

In addition, the moderate-to-strong correlation between predicted and true CD163 values (Spearman $\rho$ = 0.61, p < 0.001) supports the biological relevance of the H\&E-derived estimates.

\begin{figure}[!htbp]
  \centering
  \includegraphics[width=\linewidth]{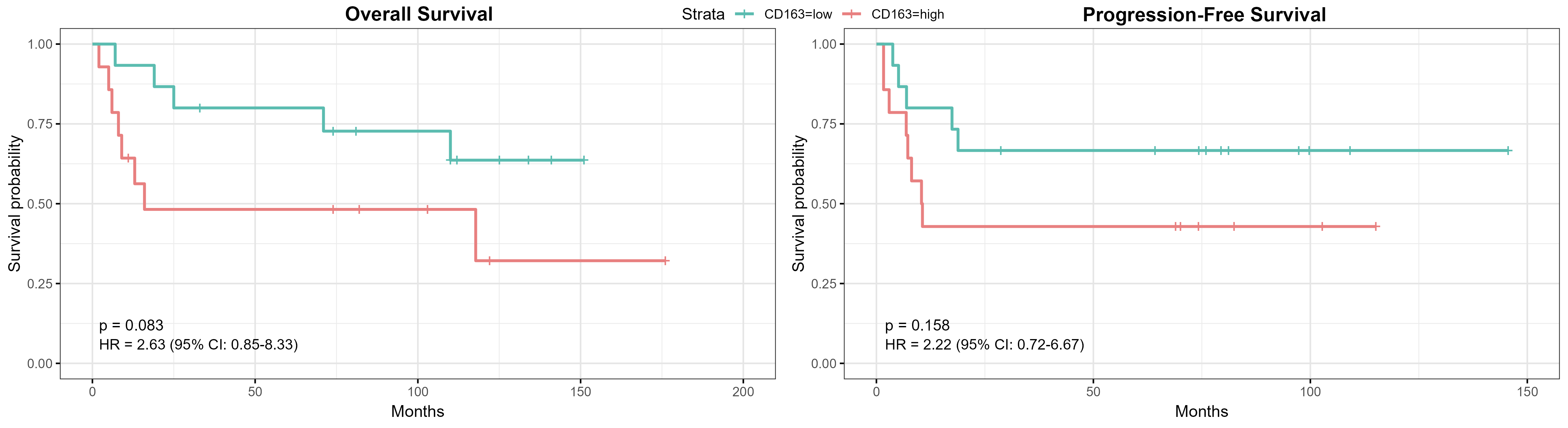}
  \caption{Association of Swin-U-Net predicted CD163 quantification expression with overall and progression-free survival. Kaplan–Meier curves showing overall survival and progression-free survival in patients stratified according to CD163 expression levels. Patients were classified into CD163 -high (red) and CD163 -low (blue) groups. Survival probability is shown on the y-axis, and time in months is shown on the x-axis.} \label{fig5}
\end{figure}

Patients with high CD163 expression had a worse overall survival, with a hazard ratio of 2.63 (95\% CI: 0.85–8.33; p = 0.083 by log-rank) and shorter progression-free survival, with a hazard ratio of 2.22 (95\% CI: 0.72–6.67; p = 0.158 by log-rank). Neither of these differences reached statistical significance Figure \ref{fig5}. The present results are therefore interpreted as a proof-of-concept demonstrating a clinically coherent trend rather than a validated clinical surrogate. To assess the clinical validity of the deep learning predictions, two pre-specified formal concordance approaches were implemented.

The IHC-derived prognostically validated threshold of 20.04\% was applied directly to the continuous CD163 area-fraction estimates produced by Swin-U-Net predictions of CD163 , and the resulting binary classification was compared to the IHC reference. 
Swin-U-Net achieved a concordance of 90.4\% (Cohen's $\kappa = 0.813$) at this 
threshold, with a sensitivity of 92.9\% and a specificity of 89.5\% ($\text{FP} = 1$, 
$\text{FN} = 1$). Continuous-level agreement between Swin-U-Net predicted area 
fractions and IHC reference values was high (linear regression: $\alpha = 1.948$, 
$\beta = 1.06$, $R^2 = 0.883$, Adj.\ $R^2 = 0.881$, $p < 0.001$), confirming that 
DL-predicted area fractions track the IHC signal across the full dynamic range.

\section*{Discussion}

In this study, we demonstrated that deep learning-based analysis of standard H\&E-stained sections can work as surrogate IHC-based CD163 TAMs quantification in patients with DLBCL. 
Our findings confirm that a high density of CD163 -enriched TAMs (cutoff $>$20.04\%) associated with adverse clinical outcomes in multivariate analysis, with an OS hazard ratio of 3.24 (95\% CI: 0.94--11.15; p = 0.062) and PFS hazard ratio of 2.41 (95\% CI: 0.80--7.32; p = 0.120), adjusted for IPI, molecular subtype (Hans GCB/non-GCB), EBV status, and age, and a nearly three-fold unadjusted increase in the risk of disease progression (HR 2.88) and mortality (HR 2.73).

These results align with previous reports on the immunosuppressive role in the tumor microenvironment in aggressive lymphomas~\cite{lin2023nuclei, cioroianu2019tumor} and underscore the potential of CD163 macrophage quantification in patients with DLBCL to streamline risk stratification.
The evaluation of the YOLOv11 family allowed us to draw the following insights. The finding that the Nano variant matched the performance of larger models (Medium/Extra-Large) suggests that TAM detection in H\&E is constrained more by phenotypic ambiguity and labeling noise than by raw model capacity. Despite the fact that the YOLOv11 model's metrics were higher than those of the basic U-Net model, they were still lower than those of the Cerberus and Swin-U-Net models. This was especially true in cases of high macrophage density, which is critical in real-world clinical settings.
Contrary to our initial hypothesis, the domain-specific Cerberus-ResNet34 backbone did not outperform the general-domain Swin-U-Net, although it achieved the highest Recall of 0.656. This suggests that while histopathology-specific pre-training enhances sensitivity to macrophage features, the architectural advantages of Transformers in handling ``crowded'' scenes may be more decisive than the pre-training domain alone. Among the Cerberus variants, the improved performance of U-Net++ and the optimized U-Net3+ underscores the necessity of dense skip connections. These connections mitigate the loss of spatial information during downsampling, which is vital when segmenting cells with irregular, non-spherical morphologies like macrophages
Although HoVer-Net is the de facto standard for nuclear segmentation in solid-tumour settings~\cite{zhang2024hovernet}, its performance in our DLBCL cohort (F1 $0.613 \pm 0.018$) was below that of Cerberus-U-Net3+ and Swin-U-Net. The HV-branch, which is the key innovation enabling dense nucleus separation in HoVer-Net, was excluded from this implementation because our task reduces to binary segmentation of macrophage pixels rather than full nuclear instance separation. Consequently, the model operated as a NP-branch-only segmentor, forgoing the crowding disambiguation that is HoVer-Net's primary competitive advantage.

M2-associated macrophages in H\&E-stained DLBCL sections frequently exhibit large nuclei with irregular or kidney-shaped contours and abundant eosinophilic cytoplasm. However, these features substantially overlap with activated histiocytes, tingible-body macrophages, follicular dendritic cells, and large neoplastic B-cells with clear cytoplasm, limiting reliable morphology-based identification. In our dataset, the primary source of false positives---as reflected in the relatively higher false-positive rates of U-Net and YOLOv11--is likely the misclassification of large lymphoma cells and histiocytic elements. Swin-U-Net's shifted-window self-attention appears to better leverage the spatial context of the surrounding tissue (e.g., the characteristic clustering of small lymphocytes around macrophages) to disambiguate these cell types, explaining its superior Precision and F1 despite a lower Recall.
While standard CNNs excel at local feature extraction, they often struggle with the architectural complexity of lymphoid tissue, where macrophages and neoplastic B-cells are densely packed. The hierarchical ViT encoder, leveraging shifted-window self-attention, appears better equipped to model the long-range semantic dependencies required to distinguish pleomorphic TAMs from their background. Interestingly, despite being pre-trained on natural images (ImageNet), the Swin Transformer demonstrated remarkable adaptability to the stochastic textures of histopathological slides.

In a proof-of-concept setting, our study shows that automated quantification of CD163 using DL models can partly replicate the predictive power of traditional IHC in DLBCL. Clinical validity was assessed by two independent formal approaches. A detailed case-level analysis of the two discordant patients in the Swin-U-Net Approach 2 classification (FP = 1, FN = 1) is instructive regarding the sources of residual error. In the first case, the model predicted a CD163 area fraction of 22.3\%, whereas IHC indicated 14.3\% (false positive); this patient subsequently died with an overall survival of 12 months. The model's over-estimation likely reflects false-positive classification of large neoplastic B-cells or histiocytic elements as CD163 macrophages--a known confound in dense DLBCL tissue--rather than a systematic bias. Notably, the patient's short OS is consistent with an aggressive disease course, and it cannot be excluded that the model detected a broader immunosuppressive microenvironmental signal that was not fully captured by the single-ROI IHC quantification. In the second discordant case, the IHC value was 20.2\% (above the 20.04\% threshold) whereas the model predicted 16.5\% (false negative). This under-estimation is consistent with the error pattern visible in Figure~\ref{fig4}C: when macrophages are clustered in tight aggregates, all models--and Swin-U-Net in particular--tend to miss a proportion of cells because clustered boundaries are difficult to resolve at the patch level. Both discordant cases reflect challenges that are intrinsic to H\&E-based macrophage identification rather than model-specific failures, and they motivate future work on instance-aware post-processing (e.g., marker-controlled watershed) and whole-slide inference to aggregate evidence across the full tissue section rather than a single representative ROI.
The survival stratification based on Swin-U-Net predictions shows a directionally consistent, clinically plausible trend (OS: HR 2.63, p = 0.083), non-significant due to the modest cohort size (n = 52). The present findings are therefore appropriately interpreted as a proof-of-concept demonstrating clinical feasibility, with the two concordance approaches providing the formal evidence of clinical validity requested by peer review; prospective multi-centre validation in a larger cohort is required before clinical deployment.
Inference time is a critical parameter for clinical deployment. Patch-level inference times measured on NVIDIA A100 40~GB GPU (batch size 1, single $256 \times 256$ patch) for a typical ROI of $2560 \times 1456$ pixels, this corresponds to a total processing time (patch extraction + inference + mask reconstruction) of approximately 25~seconds per patient using Swin-U-Net, which is clinically acceptable for a batch-processing scenario.

\subsection*{Limitations}

Despite the promising results, this study is subject to several limitations. First, the analysis was conducted on a relatively small cohort (n=52) from a single center. Validation on larger, multi-centric cohorts is required to confirm the robustness of these risk stratification thresholds across diverse patient populations and staining protocols. 
Second, the consecutive-section design introduces a methodological limitation that warrants explicit discussion. H\&E and IHC sections were obtained from adjacent 4--5~$\mu$m serial sections rather than the same tissue plane. While expert annotators used the IHC as a spatial reference map and the typical sub-cellular displacement at 20$\times$ magnification is unlikely to cause systematic misannotation, a formal affine or deformable slide registration step would further reduce residual spatial uncertainty. We recommend incorporating automated slide registration in future prospective studies. Annotation errors arising from H\&E--IHC displacement affect the ground-truth mask and training signal uniformly across all models, and are therefore unlikely to systematically bias the comparative ranking of architectures, though they may set a ceiling on achievable F1 scores.
Third, the use of CD163 as a single marker for M2-phenotype TAM identification is a genuine limitation of the study design, shared with a large body of published DLBCL TAM literature. CD163 is the most widely used single marker for M2-polarised macrophages in clinical practice, but it is not fully specific and can label monocyte-derived cells and some other myeloid populations. CD68/CD163 dual-staining would provide a more specific definition of true M2 macrophages (CD68+/CD163+) by excluding CD163+/CD68$^-$ non-macrophage populations. In the DLBCL lymph node context, this distinction is less pronounced than in solid tumours, as CD163+/CD68$^-$ non-macrophage populations are less prevalent; nevertheless, we cannot claim that CD163 single-marker staining confirms macrophage phenotype with certainty. This study used archival tissue from routine clinical practice, where only single-marker CD163 IHC was performed. CD68/CD163 dual-staining is recommended for future prospective validation studies. Future work will focus on scaling the dataset and optimizing the Swin-U-Net segmentation approach.

\section*{Conclusion}
Our findings demonstrate that CD163 TAM density associated with adverse outcomes in DLBCL, where a threshold of approximately 20\% distinguishes patients with significantly reduced overall and progression-free survival (unadjusted HR $>$2.7; multivariate OS HR 3.24, 95\% CI: 0.94--11.15, p = 0.062; multivariate PFS HR 2.41, 95\% CI: 0.80--7.32, p = 0.120), after adjustment for IPI, molecular subtype, and age. The failure to reach p $<$ 0.05 in the multivariate models reflects the limited present cohort (n = 52) rather than absence of independent prognostic signal; both hazard ratios are clinically substantial and directionally consistent. This underscores the clinical relevance of quantifying tumor-associated macrophages to characterise disease aggressiveness. From a computational perspective, we observed that increasing model complexity within the YOLOv11 family (from Nano to Extra-Large) did not yield proportional performance gains, identifying the YOLOv11n-seg variant as the most efficient iteration for resource-constrained environments. The inclusion of HoVer-Net, a de facto standard for nuclear segmentation in digital pathology demonstrated that domain-specific pre-training on solid-tumour data provides modest but meaningful benefit over the baseline U-Net while remaining below the performance of Cerberus+U-Net3+ transfer learning approach with a histological encoder achieved the highest Recall (0.656) and IoU score (0.471), but it was inferior to Swin-U-Net with respect to F1-score (0.633) and Precision (0.694). Additionally, Swin-U-Net showed more accurate border detection and fewer false positives compared to other methods analyzed.  Finally, this proof-of-concept study shows that automated quantification of CD163 using the Swin-U-Net model demonstrates potential as a scalable candidate surrogate for IHC-based prognostic assessment of tumor-associated macrophages in diffuse large B-cell lymphoma, warranting prospective multi-centre validation in larger cohorts.

\bibliography{sample}

\section*{Funding declaration}
The work was supported by the Russian Science Foundation grant № 25-71-10088

\section*{Author contributions statement}

Svetlana Illarionova and Joaquim Carreras conceived the experiments. 

Anastasiia Studenikina, Daniil Sulimov, Olga Filimonova, Dmitry Zvezdin and Arsenii Galimov conducted the experiments.

Anastasiia Studenikina, Svetlana Illarionova, Maxim Sharaev, Rifat Hamoudi and Ivan Tyukin analysed the results. 

All authors reviewed the manuscript.

\section*{Corresponding author}
Correspondence to Svetlana Illarionova.

\section*{Ethics declarations}

\subsection*{Ethics Approval and Consent to Participate}
This study was conducted in strict accordance with the guidelines of the Declaration of Helsinki and the ethical principles for medical research involving human participants. The study protocol was reviewed and approved by the Institutional Review Board of Tokai University (IRB20-156).

\subsection*{Clinical Trial Number}
Not applicable.

\subsection*{Competing of interest}
The authors declare no competing interests.

\section*{Data availability}
The datasets used and analysed during the current study available from the corresponding author on reasonable request.

\end{document}